\documentclass[conference]{IEEEtran}
\usepackage{cite}

\ifCLASSINFOpdf
\else
  \usepackage[dvips]{graphicx}
\fi
\usepackage{import}
\usepackage{color}

\usepackage[cmex10]{amsmath}
\usepackage{mathabx}
\usepackage{diagbox}
\usepackage[tight,footnotesize]{subfigure}

\usepackage{eso-pic}
\begin{document}
%
\title{Impact of Spectrum Sharing on the Efficiency of Faster-Than-Nyquist Signaling}


\author{\IEEEauthorblockN{Marwa El Hefnawy}
\IEEEauthorblockA{DOCOMO Euro-Labs \\
Landsberger Strasse 312\\
80687 Munich, Germany\\
Email: el\_hefnawy@docomolab-euro.com}
\and
\IEEEauthorblockN{Gerhard Kramer}
\IEEEauthorblockA{Institute for Communications Engineering\\ Technische Universit{\"a}t M{\"u}nchen\\
80333 Munich, Germany\\
Email: gerhard.kramer@tum.de}}



\AddToShipoutPicture*{\small \sffamily\raisebox{1.8cm}
{\hspace{1.8cm}\copyright2014 IEEE applies. This work has been accepted at the WCNC 2014. 
}}

\maketitle

\begin{abstract}
Capacity computations are presented for Faster-Than-Nyquist (FTN) signaling in the presence of interference from neighboring frequency bands. It is shown that Shannon's sinc pulses maximize the spectral efficiency for a multi-access channel, where spectral efficiency is defined as the sum rate in bits per second per Hertz. Comparisons using root raised cosine pulses show that the spectral efficiency decreases monotonically with the roll-off factor. At high signal-to-noise ratio, these pulses have an additive gap to capacity that increases monotonically with the roll-off factor.
\end{abstract}



%
\IEEEpeerreviewmaketitle

\section{Introduction}
Shannon \cite[Sec.~25]{shannon1948mathematical} showed that the capacity of an additive white Gaussian noise (AWGN) channel with bandwidth $W$ and average transmit power constraint $P$ is
\begin{equation}
\label{eq:awgnEq}
C = W \log_{2} \left( 1 + \frac{P}{WN_{0}} \right) \text{bits/sec} 
\end{equation}
where $N_{0}$ is the noise power spectral density. The capacity (\ref{eq:awgnEq}) is achieved by a sinc pulse and the spectral efficiency is
\begin{equation}
\label{eq:spec_awgn}
\eta = \log_{2} \left( 1 + \frac{P}{WN_{0}} \right) \text{bits/sec/Hz}. 
\end{equation}

Mazo \cite{mazo1975faster} introduced faster-than-Nyquist (FTN) signaling for sinc pulses where the pulses are modulated faster than the Nyquist rate. The resulting intersymbol interference (ISI) can be interpreted as a type of coding \cite{massey1989short} and is the same as correlative or partial response signaling \cite{lender1966correlative} such as the duobinary technique.\footnote{Alternatively, FTN may be viewed as a coded multi-level modulation where the transmitter-induced ISI enlarges the modulation set in addition to introducing memory (or coding). One should not, therefore, compare FTN with Nyquist ISI criterion signaling and the same modulation set unless the transmitter must use this particular modulation set for pulse shaping.} Mazo showed that increasing the modulation rate by up to $25\%$ does not affect the minimum Euclidean distance between the closest two signals when using binary antipodal modulation. Thus, the coding induced by FTN signaling increases the spectral efficiency at high signal-to-noise ratio (SNR). Non-orthogonal transmission schemes such as FTN \cite{FTN_marwa}, \cite{FTN_ALU} are receiving renewed attention for their potential to increase capacity. FTN may also be interesting for applications that need low cost transmitters and flexible rate adaptation.

In practice, it is difficult to approximate sinc pulses. One instead often analyzes square root raised cosine (RRC) pulses that decay more quickly than sinc pulses and can be approximated more accurately. Rusek, Anderson and \"Owall show in \cite{FTN2013} and \cite{rusek2009constrained} that FTN signaling achieves a substantially higher spectral efficiency than (\ref{eq:spec_awgn}) if the comparison is based on the 3-dB power bandwidth of RRC pulses with independent and identically distributed (i.i.d.) Gaussian symbols. The calculations are performed for a single channel, i.e., there is no interference or spectral sharing.

We revisit this comparison by viewing bandwidth as a shared resource where the spectral efficiency $\eta$ is computed by normalizing the sum rate of $K$ users (or systems) by an overall bandwidth of approximately $KB$ Hz, where $B$ is the bandwidth assigned to each user. For example, for Shannon's sinc pulse, every user receives $B=W$ Hz of non-overlapping bandwidth and the spectral efficiency is given by (\ref{eq:spec_awgn}).
One may try to improve $\eta$ by using non-orthogonal signaling such as FTN. Of course, now the users experience interference. Our main goal is to explore the spectral efficiency of FTN signaling from the shared resource perspective.

This paper is organized as follows: Section \ref{sec2} analyzes the capacity of FTN signaling for a single user, Section \ref{sec3} defines and analyzes spectral efficiency for a multiaccess channel with spectrum sharing, and Section \ref{sec4} discusses the results further for low and high SNR.

 


\section{Capacity}
\label{sec2}

The Nyquist \emph{rate} usually refers to twice the bandwidth of a bandlimited signal. The Nyquist intersymbol interference (ISI) \emph{criterion} refers to the requirement that sampling at regular intervals incurs no ISI \cite[p.~557]{proakis}. For sinc pulses, Nyquist-rate sampling satisfies the Nyquist ISI criterion, so that FTN refers to sampling faster than both the Nyquist rate {\em and} a Nyquist ISI criterion rate. However, in terms of capacity there is no need to sample faster than the Nyquist rate for linear AWGN channels \cite[Sec.~19]{shannon1948mathematical}. For pulses other than sinc pulses, on the other hand, it might be interesting to sample faster than the fastest Nyquist ISI criterion rate \cite{FTN2013}.

A FTN signal with complex pulse shape $h(t)$ is given by

\begin{equation}
\label{eq:s(t)}
S(t) = \sum\limits_{k=0}^{\infty}  B[k]h(t- k\tau  T).
\end{equation}
where the complex random symbols $B[k]$ are sent at rate $1/(\tau T)$ Hertz. The FTN rate is $1/\tau$ where $0<\tau\leq1$. We choose $h(t)$ to have unit energy, i.e., we choose
\begin{equation}
\label{eq:energy}
\int_{-\infty}^{\infty}| h(t)|^{2} dt = 1.
\end{equation}

\begin{figure}[!t]
\centering
\includegraphics[width=3in]{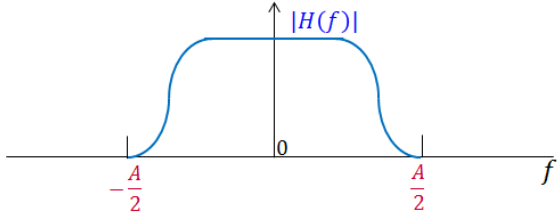}
\caption{Magnitude of the spectrum of a complex-valued signal with absolute bandwidth $A$.}
\label{SU_FTN}
\end{figure}

Suppose that $S(t)$ is constrained to have the spectrum $H(f)$ with absolute bandwidth $A$, as shown in Fig. \ref{SU_FTN}. For i.i.d. complex $B[k]$ with variance $P$, let SNR$(f)$ be the SNR at frequency $f$, i.e., define 
\begin{equation}
\text{SNR}(f) \buildrel \Delta \over = \frac{\lim_{L\to\infty} \text{E}\left[ \Gamma_{L}\right]}{N_{0}} = \frac{P|H(f)|^{2}}{N_{0}}
\end{equation}
where
\begin{equation}
\Gamma_{L} = \frac{1}{L}\left| \int_{-\infty}^{\infty} \sum_{k=0}^{L} B[k]h(t-k\tau T) e^{-j2\pi ft} dt\right| ^{2}. \nonumber
\end{equation}
Note that SNR$(f)$ does not depend on the FTN rate $1/\tau$. The capacity is achieved with proper complex Gaussian $B[k]$ and is (see \cite{rusek2009constrained} and \cite{massey93})
\begin{align}
\label{eq:cFTN}
C_{FTN} (\alpha) &= \int_{-\infty}^{\infty}\log_{2} \left( 1 + \text{SNR}(f) \right) df \nonumber\\
&= \int_{-\frac{A}{2}}^{\frac{A}{2}} \log_{2} \left( 1+\frac{P|H(f)|^{2}}{N_{0}} \right) df.  
\end{align}

\begin{figure*}[!t]
\centering
\includegraphics[width=5in]{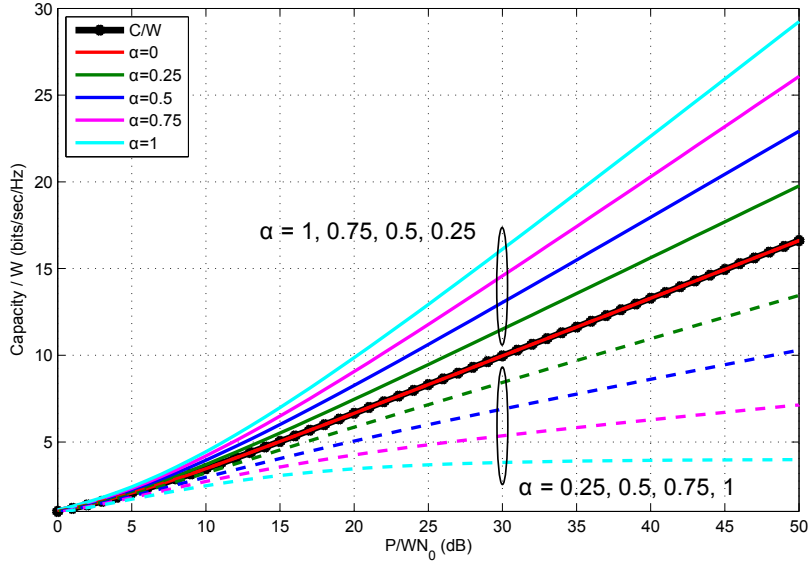}
\caption{Normalized capacities of FTN for RRC pulses and different roll-off factors $\alpha$. The solid lines represent the normalized capacities without interference and the dashed lines represent the normalized capacities (and spectral efficiencies) with interference and spectral offsets of $B=W$ Hertz.}
\label{ftn_MUSU}
\end{figure*}

For example, a RRC pulse with roll-off factor $\alpha$, $0\leq \alpha \leq 1$, has $A=(1+\alpha)W$ and

\begin{eqnarray}
\label{eq:RRC_pulse}
|H(f)|^{2} &=&
\begin{cases}
 1/W, \quad |f| \leq (1 - \alpha)\frac{W}{2} \\

 \frac{1}{2W}\left[1 + \cos\left(\frac{\pi}{\alpha W}\left[|f| - (1 - \alpha)\frac{W}{2}\right]\right)\right],\\
       \quad\quad\quad\quad (1 - \alpha)\frac{W}{2} < |f| \leq (1 + \alpha)\frac{W}{2} \\
0,
       \quad\quad\quad \mbox{else.} 
\end{cases}
\end{eqnarray}
The capacity (\ref{eq:cFTN}) is thus
\begin{align}
\label{eq:C_FTN_sec2}
C_{FTN} (\alpha) &= (1-\alpha)W \log_{2} \left( 1 + \frac{P}{WN_{0}}\right) \nonumber\\ &\quad+  2\alpha W \log_{2} \left( \frac{1+\frac{P}{2WN_{0}}+\sqrt{1+\frac{P}{WN_{0}}}}{2}\right)
\end{align}
where $ W=1/T $ and where we have used (see \cite[p.~531]{gradshteyn2000table})
\begin{equation}
\label{eq:ab}
\int_{0}^{\pi} \ln (a+b\cos x) dx = \pi \ln{\frac{a+\sqrt{a^{2}-b^{2}}}{2}}
\end{equation}
for $ a \geq |b| > 0$. The normalized capacities $C_{FTN}(\alpha)/W$ are plotted as the solid curves in Fig. \ref{ftn_MUSU} for various $\alpha$ (note that $W$ is the 3 dB bandwidth used in \cite[Sec.~3]{FTN2013}). Observe that the capacities increase with the roll-off factor due to the bandwidth expansion. The asymptotic gain in slope as compared to $C$ is proportional to the \emph{excess} bandwidth $\alpha W$ for high SNR \cite[Sec.~1]{rusek2009constrained}. To see this, note that (\ref{eq:C_FTN_sec2}) gives 
\begin{equation}
\label{eq:limit}
\lim_{\rho\to\infty} \frac{C_{FTN}(\alpha)}{\log_{2}\rho}= \left(1+\alpha\right)W
\end{equation}
where $\rho=\frac{P}{WN_{0}}$.

\section{Spectral Efficiency}
\label{sec3}

\begin{figure*}[!t]
\centering
\includegraphics[width=5in]{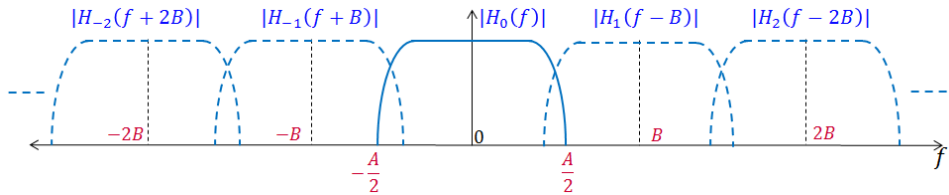}
\caption{Spectral sharing with different spectra for each user.}
\label{MU_FTN_general}
\end{figure*}

Consider a $K$-user multiaccess channel (MAC) and define the spectral efficiency as the sum-rate in bits per second per Hertz. We place the signals of the $K$ users at the center frequencies $kB, k = 0, \pm 1, \pm 2, ..., \pm (K-1)/2$, where we assume that the number of users $K$ is odd and where the centered spectrum of user $k$ is $H_{k}(f)$, see Fig. \ref{MU_FTN_general}. All spectra have at most bandwidth $A$. Note that choosing $B=0$ has all users transmitting in the same frequency band. We focus on MACs because we are interested in cellular uplinks.

Another natural scenario is a $K$-user interference network, but now the crosstalk gains play an important role in how the receivers should operate. For example, for weak interference it is usually best to treat interference as noise while for strong interference one may wish to decode the interfering signal. The $K$-user interference network becomes a compound MAC if all crosstalk gains are similar to the direct gains, and this is one further motivation for studying MACs. 

We have decided to focus on sum-rate for simplicity. One may instead wish to study weighted sum rates or other utility functions such as proportionally fair rates. These studies may lead to other conclusions, although we expect that the insight for sum-rate carries over to other important cases.

For the sum-rate, we first show that sinc pulses maximize the spectral efficiency. The total bandwidth and power are the respective

\begin{align}
\begin{split}
B_\text{tot} &= A+(K-1)B\\
P_\text{tot} &= \sum_{k=-\frac{(K-1)}{2}}^{\frac{(K-1)}{2}} P_{k}
\end{split}
\end{align}
where $P_{k}$ is the transmit power of user $k$. Suppose the average power per Hertz is constrained to be $P_\text{tot}/B_\text{tot}$. The power spectral density of the received signal is 
\begin{equation}
  N_0 + \sum_k P_k |H_k(f-kB)|^2.
\end{equation}

To upper bound the spectral efficiency, we compute
\begin{eqnarray}
\eta &\overset{\text{(a)}}\leq& \frac{1}{B_\text{tot}}\int_{-\frac{B_\text{tot}}{2}}^{\frac{B_\text{tot}}{2}}\log_{2} \left( 1+ \frac{\sum_{k}P_{k}|H_{k}(f-kB)|^2}{N_{0}}\right) df\nonumber\\
&\overset{\text{(b)}}\leq& \log_{2}\left( 1+ \frac{1}{B_\text{tot}N_{0}}\sum_{k}P_{k}\int_{-\frac{B_\text{tot}}{2}}^{\frac{B_\text{tot}}{2}}|H_{k}(f-kB)|^2 df \right)\nonumber\\
&\overset{\text{(c)}}=& \log_{2} \left(1+ \frac{P_\text{tot}}{B_\text{tot}N_{0}} \right). \label{eq:spect_proof}
\end{eqnarray}
where (a) follows by a classic maximum entropy result (see \cite{massey93}), (b) follows by Jensen's inequality, and (c) follows by (\ref{eq:energy}) and Parseval's identity. Furthermore, we achieve equality in (\ref{eq:spect_proof}) by using sinc pulses with bandwidth $A=B=W$, and choosing $P_{k}=P_\text{tot}/K$ for all $k$, see (\ref{eq:spec_awgn}). We also achieve equality by using sinc pulses with $A=W$, $B=0$, and any $P_k$ such that $\sum_k P_k = P_\text{tot}$. We note that the bound (\ref{eq:spect_proof}) is valid irrespective of the receiver algorithm. This means that sinc pulses maximize the spectral efficiency even if multi-user detection is permitted, as long as the total transmit power per Hertz is constrained to be $P_\text{tot}/B_\text{tot}$.

\begin{figure*}[!t]
\centering
\includegraphics[width=5in]{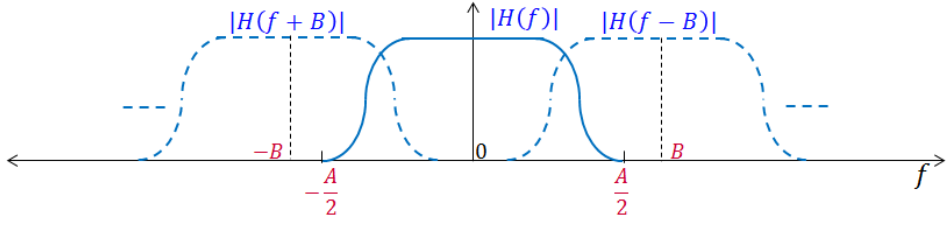}
\caption{Spectral sharing with the same spectrum for each user and $\frac{A}{2}\leq B \leq A$.}
\label{MU_FTN}
\end{figure*}

Having established the optimality of Shannon's approach, we now calculate the spectral efficiency of FTN for RRC pulses. To avoid having to consider too many cases, we make the following simplifications. We choose:
\begin{itemize}
\item to treat interference as noise, i.e., the signal-to-inteference-plus-noise-ratio (SINR) for user $k=0$ at frequency $f$ is 
\begin{equation}
\text{SINR}(f,B) = \frac{P_{0}|H_{0}(f)|^2 }{N_{0} + \sum_{k\neq 0}P_{k}|H_{k}(f-kB)|^2 };
\end{equation} 
\item to use Gaussian signaling so that we may use SINR$(f,B)$ to compute reliable communication rates, i.e., we have the information rate
\begin{align}
\label{eq:SPEC_FTN_general}
C^{'}_{FTN}(\alpha,B) = \int_{-\infty}^{\infty}\log_{2} \left( 1 + \text{SINR}(f,B) \right) df;
\end{align}
\item $H_k(f)=H(f)$ for all $k$, where $H(f)$ is the RRC pulse with $A=(1+\alpha)W$, see (\ref{eq:RRC_pulse}) and Fig. \ref{MU_FTN}; 
\item $P_k=P$ for all $k$; 
\item $A/2 \le B \le A$ so that there is at most one interferer at each frequency, i.e., we have 
\begin{equation}
\text{SINR}(f,B) = \frac{P |H(f)|^2 }{N_{0} + P |H(f-B)|^2 }
\end{equation}
where $0 \le f \le A$.
\end{itemize}

\begin{figure*}[!t]
\centering
\includegraphics[width=5in]{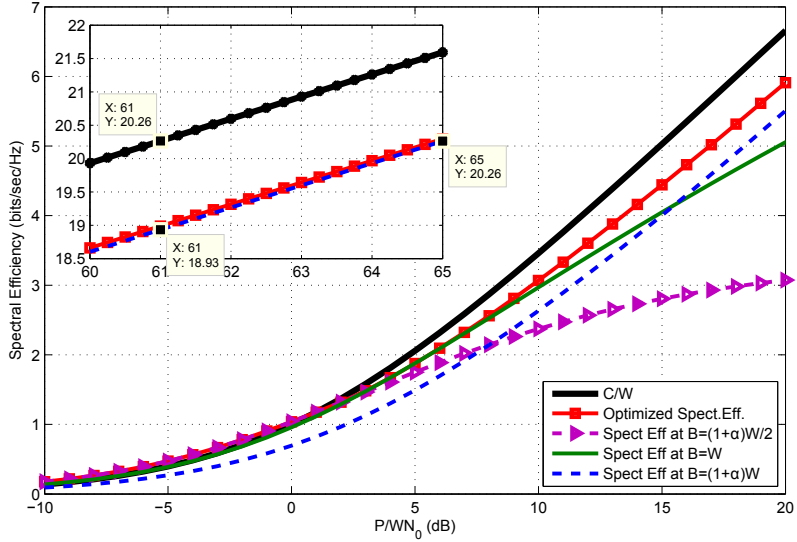}
\caption{Spectral efficiencies for roll-off factor $\alpha=0.5$.}
\label{ftn_B_1}
\end{figure*}

There are now four cases to consider for RRC pulses. 
\begin{enumerate} 
\item $B \ge (1+\alpha)W$: there is no interference;
\item $W \le B \le (1+\alpha)W$: only the cosine-portions of the RRC pulses overlap;
\item $\max[(1-\alpha)W,(1+\alpha)W/2] \le B \le W$: the cosine portions overlap the flat portions of the RRC pulses but the flat portions do not overlap;
\item $(1+\alpha)W/2 \le B \le (1-\alpha)W$: requires $0 \le \alpha \le 1/3$ and that the flat portions of the pulses overlap.
\end{enumerate}

We treat the first two cases here and the next two cases in Appendix \ref{app:A}. For $B \ge (1+\alpha)W$ there is no interference and we have (see (\ref{eq:C_FTN_sec2})) 
\begin{eqnarray}
\label{eq:B=2w(1+alpha)}
C^{'}_{FTN}(\alpha,(1+\alpha)W) = C_{FTN}(\alpha).
\end{eqnarray}
For $W\leq B\leq (1+\alpha)W$, we have 
\begin{align*}
&C^{'}_{FTN}(\alpha,B) = (1-\alpha)W \log_{2} \left( 1 + \frac{P}{WN_{0}}\right) \nonumber\\ &+2 \int_{0}^{B-W}\log_{2}  \left( 1 + \frac{P}{2WN_{0}}\left[1 + \cos\left(\frac{f\pi}{\alpha W}\right)\right]\right) df
\end{align*}
\begin{align}
&+2 \int_{B-W}^{\alpha W} \log_{2} \Bigg( 1 +\nonumber\\&  \frac{P\left[1 + \cos\left(\frac{f\pi}{\alpha W}\right)\right]}{2WN_{0}+P\left[1 + \cos\left(\frac{(f- B + (1-\alpha)W )\pi}{\alpha W}\right)\right]}\Bigg)df.
\label{eq:SPEC_FTN}
\end{align}

For the special case $B=W$, we compute (see Appendix \ref{app:B}):

\begin{align}
\label{eq:B=2w}
C^{'}_{FTN}(\alpha,W)  &= (1+\alpha)W \log_{2} \left( 1 + \frac{P}{WN_{0}}\right) \nonumber\\ &\quad\;- 2\alpha W \log_{2} \left( \frac{1 + \frac{P}{2WN_{0}} + \sqrt{1 + \frac{P}{WN_{0}}}}{2}  \right).\nonumber\\
\end{align}

\begin{figure*}[!t]
\centering
\includegraphics[width=5in]{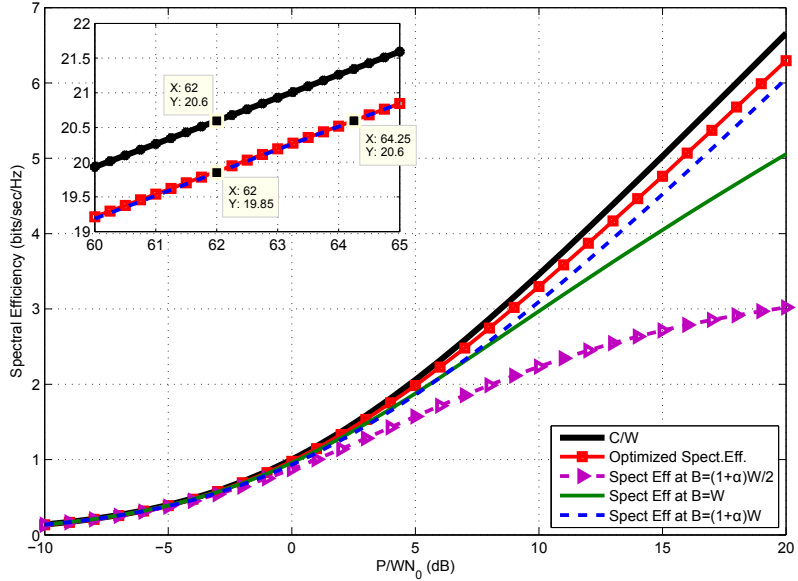}
\caption{Spectral efficiencies for roll-off factor $\alpha=0.5$ with transmit power normalized to $P/W$ Watts/Hz.}
\label{ftn_B_1_Normalized_Power}
\end{figure*}

To compute the spectral efficiency, we divide (\ref{eq:SPEC_FTN_general}) by the total bandwidth $(1+\alpha)W + (K-1)B$. As $K$ gets large, the bandwidth per user is approximately $B$ and the spectral efficiency is
\begin{equation}
\label{eq:SPEC_FTN_B}
\eta^{'}_{FTN}(\alpha,B) = \frac{1}{B} C^{'}_{FTN}(\alpha,B).
\end{equation}
For $B=W$, we plot the corresponding spectral efficiencies $\eta^{'}_{FTN}(\alpha,W)$ as the dashed curves in Fig. \ref{ftn_MUSU}. Observe that the spectral efficiency \emph{decreases} with increasing $\alpha$ which is the opposite as in \cite[Sec.~2]{FTN2013} where the interference is not accounted for. 

The spectral efficiencies optimized over $B$ satisfying $(1+\alpha)W/2 \leq B \leq (1+\alpha)W$ are shown in Fig. \ref{ftn_B_1} for the roll-off factor $\alpha=0.5$. The reader may find it strange that choosing $B=(1+\alpha)W/2$ beats Shannon's $C/W$ curve at low SNR. We discuss the low and high SNR effects next.

\section{Low and High SNR}
\label{sec4}

At low SNR or $\frac{P}{WN_{0}}\rightarrow 0$, the spectral efficiency (\ref{eq:SPEC_FTN_B}) can be approximated as 
\begin{equation}
\label{eq:lowSNR}
\eta^{'}_{FTN}(\alpha,B) \approx \frac{P}{BN_{0}} \log_{2} e
\end{equation}
and it is best to choose $B$ as small as possible. In fact as $B\rightarrow 0$ the approximation (\ref{eq:lowSNR}) remains valid and we can achieve an arbitrarily large spectral efficiency. This perhaps unexpected behavior is because the transmit power per Hertz for large $K$ is $P_{\text{tot}}/B_{\text{tot}} \approx P/B$, i.e., the power per Hertz increases as $B$ decreases. In comparison, the transmit power per Hertz for orthogonal transmission with Shannon's sinc pulses is $P/W$. We should thus normalize (\ref{eq:lowSNR}) by multiplying by $B/W$, and we arrive at the same spectral efficiency for all positive $B$. The result (\ref{eq:lowSNR}) remains valid for $B>A$ also, and this relates to the optimality of bursty signaling at low SNR.

This observation also explains the low-SNR behavior of the curves in Fig. \ref{ftn_B_1}: the gains and losses for $B\neq W$ as compared to $B=W$ are because the transmit power per Hertz $P/B$ depends on $B$. If we normalize to $P/W$ Watts/Hz and then optimize over $B$ in the range $(1+\alpha)W/2 \leq B \leq (1+\alpha)W$ we arrive at the curves shown in Fig. \ref{ftn_B_1_Normalized_Power}. Now any choice for $B$ gives the same spectral efficiency at low SNR, as expected.

At high SNR or $\frac{P}{WN_{0}}\rightarrow \infty$, the spectral efficiency (\ref{eq:SPEC_FTN_B}) based on (\ref{eq:SPEC_FTN}) can be approximated as  
\begin{equation}
\label{eq:highSNR}
\eta^{'}_{FTN}(\alpha,B) \approx \frac{2B-(1+\alpha)W}{B}\log_{2}\left( \frac{P}{WN_{0}} \right) 
\end{equation}
and it is best to choose $B$ as large as possible, i.e., $B=(1+\alpha)W$ which corresponds to no interefernce. The resulting spectral efficiency pre-log is 1. The spectral efficiency pre-log for $B=W$ and high SNR is $1-\alpha$, which is the high-SNR slope of the dashed curves in Fig. \ref{ftn_MUSU}.

\begin{figure}[!t]
\centering
\includegraphics[width=3in]{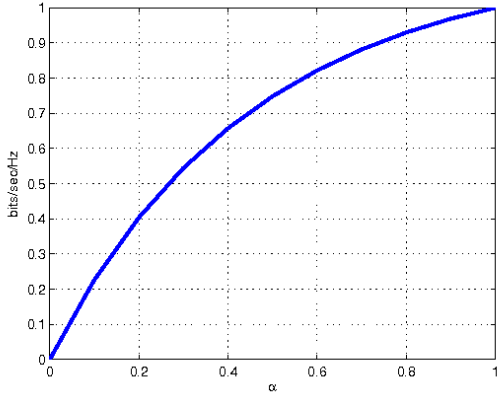}
\caption{Additive gap of the spectral efficiency at high SNR.}
\label{alpha_loss}
\end{figure}

Finally, a more precise version of (\ref{eq:SPEC_FTN_B}) for $B=(1+\alpha)W$ and high SNR gives 
\begin{equation}
\eta^{'}_{FTN}(\alpha,B) \approx \log_{2} \left(\frac{P}{WN_{0}}\right)-\frac{4\alpha}{1+\alpha} \quad \text{bits/sec/Hz}.
\end{equation}
Note that there is an additive gap as compared to (\ref{eq:highSNR}) and this gap increases monotonically with $\alpha$. For example, the gap for $\alpha=0.5$ is 4/3 bit/sec/Hz which corresponds to a 4.01 dB loss in energy efficiency. This gap can be seen at high SNR in Fig. \ref{ftn_B_1}. The gap for $\alpha=1$ is 2 bit/sec/Hz which corresponds to a 6 dB loss in energy efficiency. After normalizing the transmit power per Hertz to $P/W$, the gap reduces to 
\begin{equation}
\frac{4\alpha}{1+\alpha} -\log_{2}(1+\alpha) \quad \text{bits/sec/Hz}
\end{equation}
which we plot in Fig. \ref{alpha_loss}. For $\alpha=0.5$ the gap is 0.75 bit/sec/Hz, i.e., the loss is 2.25 dB which can be seen at high SNR in Fig. \ref{ftn_B_1_Normalized_Power}. The gap for $\alpha=1$ is 1 bit/sec/Hz, i.e., the loss is 3 dB.

\section{Conclusion}

Spectral efficiency is usually considered in the context of spectrum sharing.
We showed that the spectral efficiency of RRC pulses with FTN decreases monotonically with the roll-off factor. This means that Shannon's sinc pulses are the best RRC pulses, and they are in fact the best pulses in general. At low SNR, FTN neither improves nor degrades the spectral efficiency. At high SNR, it is best to avoid interference for the models considered here.

\appendices
\section{}
\label{app:A}
For $\max[(1-\alpha)W,(1+\alpha)W/2]\leq B\leq W$, we have
\begin{align*}
&C^{'}_{FTN}(\alpha,B) = \left[2B-(1+\alpha)W\right] \log_{2} \left( 1 + \frac{P}{WN_{0}}\right) \nonumber\\ &+2 \int_{0}^{W-B}\log_{2} \Bigg(1+ \nonumber\\ &\quad\frac{2P}{2WN_{0}+P\left[1+\cos\left(\frac{(f-\alpha W)\pi}{\alpha W}\right)\right]}\Bigg) df\nonumber\\
&+ 2 \int_{W-B}^{\alpha W}\log_{2}\Bigg( 1+\nonumber\\ &\quad \frac{P\left[ 1+ \cos \left( \frac{(f+B-W)\pi}{\alpha W} \right)\right]}{2WN_{0}+P\left[ 1+ \cos\left( \frac{(f-\alpha W)\pi}{\alpha W} \right)\right]} \Bigg) df\nonumber\\
&+ 2 \int_{\alpha W}^{(1+\alpha)W-B}\log_{2}\Bigg(1+ \frac{P\left[ 1+ \cos\left( \frac{(f+B-W)\pi}{\alpha W}\right) \right]}{2WN_{0}+2P} \Bigg) df.
\end{align*}

For $(1+\alpha)W/2 \le B \le (1-\alpha)W$, we have 
\begin{align*}
&C^{'}_{FTN}(\alpha,B) = \left[2B-(1+\alpha)W\right] \log_{2} \left( 1 + \frac{P}{WN_{0}}\right) \nonumber\\ &+2 \int_{0}^{\alpha W}\log_{2} \Bigg(1+ \frac{2P}{2WN_{0}+P\left[1+\cos\left(\frac{(f-\alpha W)\pi}{\alpha W}\right)\right]}\Bigg) df\nonumber
\end{align*}
\begin{align*}
&+ 2 \left( \left(1-\alpha\right)W - B\right) \log_{2}\left( 1+ \frac{P}{WN_{0}+P}\right)\nonumber\\
&+ 2 \int_{W-B}^{(1+\alpha)W-B}\log_{2}\Bigg(1+  \frac{P\left[ 1+ \cos\left( \frac{(f+B-W)\pi}{\alpha W}\right) \right]}{2WN_{0}+2P} \Bigg) df.
\end{align*}

\section{}
\label{app:B}

We compute (\ref{eq:B=2w}) as follows:
\begin{equation*}
C^{'}_{FTN}(\alpha,B) = (1-\alpha)W \log_{2} \left( 1 + \frac{P}{WN_{0}}\right) + X_{1}
\end{equation*}

where $X_{1}$ is
\begin{align*}
&2 \int_{0}^{\alpha W} \log_{2} \left( 1 + \frac{P\left[1 + \cos\left(\frac{f\pi}{\alpha W}\right)\right]}{2WN_{0}+P\left[1 + \cos\left(\frac{(f-\alpha W)\pi}{\alpha W}\right)\right]}\right) df \\
&= 2\int_{0}^{\alpha W} \log_{2} \left( \frac{WN_{0}+P}{WN_{0} + \frac{P}{2}\left[ 1- \cos \left( \frac{f\pi}{\alpha W}\right)\right]}\right) df\\
&\overset{\text{(a)}}= 2\alpha W \log_{2} \left( 1+ \frac{P}{WN_{0}}\right) \\&\quad- \frac{2\alpha W}{\pi} \int_{0}^{\pi} \log_{2} \left(  \left( 1 + \frac{P}{2WN_{0}} \right) - \frac{P\cos(x)}{2WN_{0}} \right) dx \\
&\overset{\text{(b)}}= 2\alpha W \log_{2} \left( 1+ \frac{P}{WN_{0}}\right) \\ &\quad- 2\alpha W \log_{2} \left( \frac{1 + \frac{P}{2WN_{0}} + \sqrt{1 + \frac{P}{WN_{0}}}}{2}  \right)
\end{align*}
where (a) follows by subsitiuting $x=\frac{f\pi}{\alpha W}$ and (b) follows by (\ref{eq:ab}).

\section*{Acknowledgment}
This work was performed in the framework of the FP7 project ICT-317669 METIS. G. Kramer was supported by an Alexander von Humboldt Professorship endowed by the German Federal Ministry of Education and Research. 
%
%



\bibliographystyle{IEEEtran}
%
%
%


\end{document}